\documentclass[conference, 10pt]{IEEEtran}

\usepackage{amsmath,dsfont,bbm,epsfig,amssymb,amsfonts,amstext,verbatim,amsopn,cite}
\usepackage{subfigure,multirow,multicol,lipsum,xfrac}
\usepackage{amsthm,ulem}
\usepackage{mathtools,amsthm}
\usepackage{perpage}
\usepackage{balance}
\usepackage{url}
\usepackage{amsfonts}
\usepackage{epsfig}
\usepackage[font={small}]{caption}

\usepackage{amsfonts,amsthm,xcolor,bbm}
\usepackage{bookmark}
\usepackage{tikz}
\usetikzlibrary{angles,quotes,arrows,decorations.pathmorphing}

\def\expect{\mathop{\mbox{$\mathsf{E}$}}}

\usepackage{etoolbox}
\usepackage{algorithmicx}
\usepackage{algpseudocode}
\usepackage{pifont}
\usepackage[utf8]{inputenc}
\usepackage[T1]{fontenc}  
\usepackage[nolist]{acronym}
\MakePerPage{footnote}
\usepackage{textcomp}
\usepackage{paralist}
\usepackage[shortlabels]{enumitem}
\usepackage{bbm}
\usepackage[process=auto]{pstool}

\newcommand{\setR}{\mathbb{R}}

\newcommand{\setZ}{\mathbb{Z}}

\newcommand{\bz}{{\boldsymbol{z}}}

\newcommand{\br}{{\mathbf{r}}}

\newcommand{\bi}{{\boldsymbol{\hat\imath}}}

\newcommand{\bc}{{\boldsymbol{c}}}
\newcommand{\bch}{{\boldsymbol{\hat c}}}

\newcommand{\btc}{{\boldsymbol{\tilde c}}}
\newcommand{\btr}{{\boldsymbol{\tilde r}}}
\newcommand{\tr}{{\tilde r}}

\newcommand{\bn}{{\boldsymbol{n}}}

\renewcommand{\emph}[1]{\textit{#1}}
\newtheorem{theorem}{Theorem}

\begin{document}
\title{Soft Interference Cancellation for Random Coding in Massive Gaussian Multiple-Access}
\author{
\IEEEauthorblockN{
Ralf R. M\"uller\IEEEauthorrefmark{1}
}
\IEEEauthorblockA{
\IEEEauthorrefmark{1}Institute for Digital Communications, Friedrich-Alexander Universit\"at Erlangen-N\"urnberg, Germany\\
ralf.r.mueller@fau.de
}
}
%
%
\IEEEoverridecommandlockouts

\maketitle

\begin{acronym}
\acro{oas}[OAS]{oversampled adaptive sensing}
\acro{csi}[CSI]{channel state information}
\acro{awgn}[AWGN]{additive white Gaussian noise}
\acro{iid}[i.i.d.]{independent and identically distributed}
\acro{rhs}[r.h.s.]{right hand side}
\acro{lhs}[l.h.s.]{left hand side}
\acro{wrt}[w.r.t.]{with respect to}
\acro{rs}[RS]{replica symmetry}
\acro{rsb}[RSB]{replica symmetry breaking}
\acro{mse}[MSE]{mean squared error}
\acro{mmse}[MMSE]{minimum MSE}
\acro{sinr}[SINR]{signal to interference and noise ratio}
\acro{mf}[MF]{matched filtering}
\end{acronym}

\begin{abstract}
In 2017, Polyanskiy \cite{polyanskiy:17} showed that the trade-off between power and bandwidth efficiency for massive Gaussian random access is governed by two fundamentally different regimes: low power and high power.
For both regimes, tight performance bounds were found by Zadik et al.\ \cite{zadik:19}, in 2019.

This work utilizes recent results on the exact block error probability of Gaussian random codes in additive white Gaussian noise to propose practical methods based on iterative soft decoding to closely approach the bounds in \cite{zadik:19}.
In the low power regime, this work finds that orthogonal random codes can be applied directly. In the high power regime, a more sophisticated effort is needed.
This work shows that power-profile optimization by means of linear programming as pioneered by Caire et al.\ \cite{caire:01b}, in 2001, is a promising strategy to apply.

The proposed combination of orthogonal random coding and iterative soft decoding even outperforms the existence bounds of Zadik et al.\ \cite{zadik:19} in the low power regime and is very close to the non-existence bounds for message lengths around 100 and above.

Finally, the approach of power optimization by linear programming proposed for the high power regime is found to benefit from power imbalances due to fading which makes is even more attractive for typical mobile radio channels.
\end{abstract}

\IEEEpeerreviewmaketitle

\section{Introduction}
\label{sec:intro}
Massive multiple-access is a key component of the upcoming internet-of-things.
In contrast to classical settings, the number of users typically exceeds the number of bits an individual user aims to communicate.
Therefore, it makes sense to consider different asymptotics for massive multiple-access: Keep the message length fixed, but let the number of users grow over all bounds. This is in contrast to the classical setting in information theory where the message length becomes infinitely large, but the number of users remains constant.

This new asymptotic setting was first discussed in \cite{chen:17} and further developed by \cite{polyanskiy:17} for static, non-faded channels. A key observation of \cite{polyanskiy:17} was that a new definition of error probability is appropriate: It is sufficient if most users are able to decode their messages correctly. Thus, we refer to the per-user probability of error in the sequel, even if this is not stated explicitly.

A similar asymptotic setting, focusing on bit error probability and convolutional codes concatenated with random spreading, was first analyzed in \cite{caire:01b}, see also \cite{caire:04}. Qualitatively similar conclusions as in \cite{polyanskiy:17} were reported:
The spectral efficiency grows without need for larger energy per bit up to some limit.
Only beyond that limit, additional energy is required to further increase spectral efficiency. So, the behavior fundamentally differs in the low and the high power regime.

The existence bounds found in \cite{polyanskiy:17} were improved in the subsequent work \cite{zadik:19} which managed to very tightly quantify the tradeoff between spectral and power efficiency in the regime of high signal-to-noise ratio (SNR). For low SNR, the gap between the two bounds has remained significant. Furthermore, the bounds in \cite{polyanskiy:17} and \cite{zadik:19} were obtained by non-constructive means, i.e.\ just as Shannon's 1948 random coding argument, they do not hint  towards any algorithm that is capable to achieve them closely, in practice.

Therefore, this work intends to pursue the following three aims:
\begin{enumerate}
\item
Improve the theoretical bounds in \cite{zadik:19}.
\item
Propose coding and decoding schemes with polynomial complexity that closely approach the performance promised by these bounds.
\item
Investigate in which way these results for static channels carry over to fading channels.
\end{enumerate}

In order to achieve these goals, the following methods are combined, successfully:
\begin{enumerate}[A)]
\item Iterative soft cancellation of interference, i.e.\ only an attenuated version of the estimated interference is subtracted from the receive signal to reduce the potentially harmful effect of error propagation \cite{nelson:94,wang:99}. 
\item Recent results by the author to calculate the exact ensemble-averaged block-error probability of independent identically distributed (iid) Gaussian random codes \cite{mueller:20d}.
\item The optimality of orthogonal constellations with respect to block error probability.
\item
Finding the fixed-point of the iterations by tracking the evolution of multiuser efficiency of all users as pioneered in \cite{boutros:01}.
\item
Power profile optimization by linear programming as proposed in \cite{caire:01b,caire:04} to cope with the high power regime.
\end{enumerate}

In order to achieve the three aims stated above, the paper is organized as follows:
The system model and iterative soft interference cancellation, i.e.\ Method A, is introduced in Section~\ref{rac}. 
Section~\ref{lul} finds the infinite user limit for the ensemble averaged posterior block error probability of Gaussian random coding at fixed message length for a given amount of residual interference utilizing Method B.
Section~\ref{intcan} finds upper and lower bounds on the amount of residual interference after soft cancellation utilizing the Results of Section~\ref{lul} and tracks their evolution with Method C.
Since all results so far are asymptotic in nature, Section~\ref{finiteM} addresses issues that arise for a finite number of users.
Section~\ref{imcon} utilizes Method D to improve convergence of the iterations and the tightness of the bounds in the high power regime.
Section~\ref{nearfar} addresses the influence of fading and shows that it is actually helpful in the high power regime. Section~\ref{numres} discusses numerical results and Section~\ref{conc} outlines conclusions and implications.

\section{System Model}
\label{rac}
Let there be $M$ users with codewords $\bc_1,\dots,\bc_M$ that want to communicate over the Gaussian multiple-access channel
\begin{equation}
\br = \sum\limits_{m=1}^M \bc_m+ \bn
\label{sysmod}
\end{equation}
with additive white Gaussian noise (AWGN) $\bn$ of unit covariance, i.e., $\expect \bn\bn^{\dagger} = \mbox{\bf I}$.
Every user wants to transmit $K$ information bits and encodes them into the codeword $\bc_m\in\setR^{MN}$ for some $N$ such that $MN\in\setZ$.
The codeword $\bc_m$ is chosen from the set ${\cal C}_m$ of $2^{K}$ jointly independent identically distributed (iid.) Gaussian codewords by a bijective mapping to the information bits of user $m$.
The codebooks of different users are chosen statistically independent from each other.

Let the total set of all users be decomposed into a finite number of disjoint groups ${\cal G}_1,\dots, {\cal G}_J$. Within group ${\cal G}_j$, the power of every user is given by $P_j/M$, i.e., $\expect \bc_m\bc_m^\dagger = P_j\mbox{\bf I}/M$.
The powers of the users are equal within each group, but differ from group to group.
The fraction of users in group ${\cal G}_j$ is denoted by $\alpha_j = |{\cal G}_j|/M$.
The aggregate power of all users is denoted by
\begin{equation}
P=\sum\limits_{j=1}^J \alpha_jP_j.
\label{totP}
\end{equation}
The users are solely grouped to improve the convergence of successive cancellation by means of power control in Section~\ref{imcon}; see \cite{caire:04} for detailed reasons on this user grouping. All users transmit independently from any other user in the same or a different group.

Let
\begin{equation}
R= \frac{K}{N}
\label{defspeceff}
\end{equation}
denote the aggregate rate of all users.
It is sometimes referred to as spectral efficiency.
The meaning of the variable $N$ is not intuitively clear. In fact, it is a free parameter for system design. In the single user case ($M=1$), it is the blocklength of the code. In \cite{zadik:19}, its reciprocal $1/N$ is called {\it user density}.

Let all users use parallel successive decoding in an iterative manner.
That means all users are decoded in parallel resulting in estimated codewords $\bch_m$. Then, the interference is estimated for all users and cancelled from the received signal, before all users are decoded again with (hopefully) lower error probability than initially.
For any user $m$, the new estimate at iteration $i+1$ is formed from the estimate at iteration $i$ by
\begin{equation}
\bch_m^{(i+1)} = f\left(\br + p_m^{(i)} \bch_m^{(i)} - \sum\limits_{m=1}^M p_m^{(i)} \bch_m^{(i)}\right)
\end{equation}
for some decoding function $f(\cdot)$ and some soft-cancellation coefficient $p_m^{(i)}$ to be specified later on.
This process is repeated until a steady state is reached.

The signals of other users initially fully interfere.
After some iterations, only a certain fraction $v_j$ of the interference power, which group ${\cal G}_j$ had initially contributed, remains due to partially successful cancellation of interference.
At this point, the aggregate power of interference and noise is given as
\begin{equation}
\label{defI}
I = 1+\sum\limits_{j=1}^J \alpha_jv_j P_j  
\end{equation}
in the large user limit $M\to\infty$, as the power of the user of interest vanishes.

\section{Asymptotic Block Error Probability}
\label{lul}

Given a certain fraction of remaining interference, we want to calculate the posterior (conditional) block error probability of the decoder averaged over the random code ensemble in the large user limit $M\to\infty$. We will need this block error probability in Section~\ref{intcan} to find the fixed-point of the iterative cancellation process. 

We start with the unconditional block error probability which is calculated in Appendix~\ref{appproof} utilizing recent results of \cite{mueller:20d}.
\begin{theorem}
Given the Gaussian multiple-access channel defined in \eqref{sysmod} and residual interference treated as AWGN, the unconditional block error probability of any user in group ${\cal G}_j$ averaged over the random code ensemble of this same user converges almost surely to
\begin{align}
p_j &= 1- \int\limits_{\setR} {\text Q}\left(x - \sqrt{ \eta NP_j} \right) ^{2^{K}-1} {\rm D}x
\label{subf73}
\end{align}
for $M\to\infty$ with ${\rm D}x := {\text e}^{-{x^2}/2}/\sqrt{2\pi}{\rm d}x$ denoting the Gaussian measure and 
\begin{equation}
\label{defeta}
\eta = \frac 1{1+\sum\limits_{j=1}^J \alpha_jv_j P_j}
\end{equation}
denoting the multiuser efficiency \cite{verdu:98}.
\end{theorem}

The unconditional block error probability \eqref{subf73} is the symbol error probability of a $2^{K}$-dimensional orthogonal constellation in AWGN and can already be found in \cite{balakrishnan:62}, see also \cite[(5.2-21)]{proakis:00}. 
All codewords of all users are asymptotically pairwise orthogonal to each other in the large user limit.
This is a special case of a stronger result in \cite{jiang:04}: Let there be $n$ iid zero-mean Gaussian random vectors in $\beta n$ dimensions with $0<\beta<\infty$. Let $\alpha$ be the cosine of the smallest angle between any pair of them.  Then, $\alpha \sqrt{n/\ln n}$ converges almost surely to $2$, as $n\to\infty$. Note, however, that asymptotic pairwise orthogonality does not imply that codewords do not interfere with each other. Even if the interference due to the codeword of an individual user vanishes, the aggregate interference of infinitely many users may be strictly positive.

The asymptotic orthogonality allows us to even calculate some posterior block error probabilities in the large user limit.
Consider an alternative Cartesian coordinate system in $2^K$ dimensions that results from the original coordinate system by the following 2-step procedure:
\begin{enumerate}
\item
an orthonormal  transformation such that $\btc_k$, denoting the $k^{\text{th}}$ codeword of the codebook of the user of interest, is a positive multiple of the $k^{\text{th}}$ unit vector,
\item
the removal of all coordinates with index greater than $2^K$.
\end{enumerate}
The orthonormal transformation ensures that the statistical properties of all signals are preserved.
The dropped coordinates do not contain useful information about the data of the user of interest.

Let the $\btr= [\tr_1,\dots,\tr_{2^K}]$ denote the received vector in the new coordinate system.
The tildes serve to distinguish the original coordinate system in $MN$ dimensions from this newly introduced one in $2^K$ dimensions. Assume that codeword $\btc_1$ has been sent and define
\begin{align}
\tr_{k:} = \max \{\tr_k, \tr_{k+1},\dots,\tr_{2^K}\}.
\end{align}
Note that $\tr_1$ and $\tr_{2:}$ are statistically independent and $\tr_{1:} = \max \{\tr_1,\tr_{2:}\}$.
With these definitions, a decoding error occurs, if $\tr_{2:} > \tr_1$.
Conditioning on the largest component of the receive word $\tr_{1:}$, we get the posterior block error probability
\begin{align}
\label{eq23}
p_{j|\tr_{1:}} &=
\Pr( \tr_1 <\tr_{2:}| \tr_{1:}) \\
& = \frac{\int\limits_{\setR}  {\text P}_{\tr_{1}}(\tr_{2:}) \,{\text p}_{\tr_{2:}}\!(\tr_{2:}) \delta(\tr_{2:}-\tr_{1:})\,{\text d}\tr_{2:}}
{{\text p}_{\tr_{1:}}\!(\tr_{1:})}\\
&= \frac{{\text P}_{\tr_{1}}(\tr_{1:}) \,{\text p}_{\tr_{2:}}\!(\tr_{1:})}{{\text p}_{\tr_{1:}}\!(\tr_{1:})}
\end{align}
utilizing Bayes' law with ${\text P}_a(\cdot)$ and ${\text p}_a(\cdot)$ denoting cumulative distribution function and probability density function of $a$, respectively. The Dirac function $\delta(\cdot)$ occurs, since $\tr_1 <\tr_{2:}$ implies $\tr_{2:}=\tr_{1:}$.
Furthermore, exchanging random variables $\tr_1$ and $\tr_{2:}$ in \eqref{eq23} gives the probability of the complementary event.
Thus, we have
\begin{equation}
{\text p}_{\tr_{1:}}\!(\tr_{1:}) = {\text P}_{\tr_{2:}}(\tr_{1:}) \,{\text p}_{\tr_{1}}\!(\tr_{1:}) + {\text P}_{\tr_{1}}\!(\tr_{1:}) \,{\text p}_{\tr_{2:}}\!(\tr_{1:}).
\end{equation}
which leads to
\begin{align}
p_{j|\tr_{1:}}  & = \frac{1}{1 + \frac{{\text P}_{\tr_{2:}}\!(\tr_{1:}) \,{\text p}_{\tr_{1}}\!(\tr_{1:})}{{\text P}_{\tr_{1}}\!(\tr_{1:}) \,{\text p}_{\tr_{2:}}\!(\tr_{1:})} }\\
& = \frac1{1+\text{F}\left(\tr_{1:}\big/\sqrt I\right)}
\label{posterior}
\end{align}
with implicit definition of ${\text F}(\cdot)$.
Note that
\begin{align}
\tr_1 = \sqrt{NP_j}+\zeta
\end{align}
and
\begin{align}
{\text P}_{\tr_1}(x) &= {\text Q}\left ( \left(\sqrt{NP_j}-x\right)\big/\sqrt I \right).
\label{cdfrtilde1}
\end{align}
Furthermore,
\begin{align}
  p_{j|\zeta} = 1-  {\text P}_{\tr_{2:}}(\tr_1)
\end{align}
together with \eqref{pjzeta} implies
\begin{align}
{\text P}_{\tr_{2:}}(x) &= {\text Q}\left ( -x\big/\sqrt I \right)^{2^{K}-1}.
\label{cdfrtilde2}
\end{align}
With \eqref{cdfrtilde1} and \eqref{cdfrtilde2} and their derivatives, the posterior block error probability \eqref{posterior} can be evaluated for any observation $\tr_{1:}$.
In particular, we find
\begin{equation}
{\text F}(x) = \frac{
{\text Q}\left ( - x \right)
{\text e} ^{-\frac12\left(x-\sqrt{\eta NP_j}\right)^2}
}
{
{\text Q}\left ( -x+\sqrt{\eta NP_j} \right)
{\text e}^{-\frac{x^2}{2}}
(2^K-1)
}.
\end{equation}

\section{Evolution of Residual Interference}
\label{intcan}

In order to track the block error probability during iterations, we need to connect the fraction of remaining interference $v_j$ to the error probability at the previous iteration. This section serves exactly that purpose.

The remaining interference is determined by the way potential interference is cancelled.
There are various ways of performing soft interference cancellation.
Irrespective of the precise algorithm, the dynamics of the iterations can be studied by tracking the multiuser efficiency, as proposed in  \cite{boutros:01}.
The advantage of tracking multiuser efficiency in comparison to, e.g., using extrinsic information transfer charts (see \cite{richardson:08} for details), is the fact that the multiuser efficiency of all user is equal in the large system limit \cite[Proposition~2]{boutros:01}. So only a single parameter needs to be tracked.

With \eqref{defeta}, we have
\begin{align}
\eta^{(i)} & = \frac 1{1+\sum\limits_{j=1}^J \alpha_j v_j^{(i)} P_j}
\label{etadyn}
\end{align}
with $\eta^{(i)}$ and $v^{(i)}_j$ denoting the multiuser efficiency and the remaining fraction of interference in group ${\cal G}_j$, both at iteration $i$.
The goal of this section is to characterize the mapping
\begin{align}
\eta^{(i)} & \mapsto  [v^{(i+1)}_1, \dots, v^{(i+1)}_J]
\label{etavmap}
\end{align}
in order to track the evolution of the multiuser efficiency.

During iterations, the interference may become correlated to the true data. This is a severe issue, since decision rules based on Euclidean distance, as used in this work, require the statistical independence between data and interference. However, the correlation between data and interference vanishes in the large user limit, i.e., $M\to\infty$, due to the following property of the random code construction:
The codewords of all users are chosen statistically independent. Thus, they are orthogonal in the large user limit. This means that a wrong decision in iteration $i$, by means of erroneous cancellation, does not lead to an additional interference in iteration $i+1$ that points into the same direction as the true signal, as it would be the case for, e.g., binary antipodal constellations. In contrast, it creates  additional interference that is orthogonal to the true data. 

For the calculation of error probability, we rely on self-ergodicity.
Self ergodicity means that in an infinite population of independent users, the relative frequency of decoding errors matches its statistical distribution. Thus, the instantaneous interference power after interference cancellation based on potentially erroneous decoding also equals its statistical expectation.


If we have received word $\br$ and decided for a codeword $\bch_m$, this decision is correct with probability $1-p_{j|\br}$ for all $m\in{\cal G}_j$.
Paying tribute to potentially wrong decisions, we do not fully subtract the codeword $\bch_m$ from the received word $\br$, but only subtract $q_{j|\br}\, \bch_m$ with some soft-cancellation factor $0\le q_{j|\br} \le 1$ depending on the conditional error probability $p_{j|\br}$.
After soft cancellation, the remaining interference power due to any user in group ${\cal G}_j$ is
\begin{equation}
\label{intpow}
\left[(1-q_{j|\br})^2 (1-p_{j|\br}) + \left(1+q_{j|\br}^2\right) p_{j|\br}\right] \frac{P_j}M
\end{equation}
on average.
Note again that all codewords are statistically independent. In case of erroneous cancellation, the interference does not add in amplitude, but in power.
Direct optimization of \eqref{intpow} leads to the soft-cancellation rule
\begin{equation}
\label{softcanrule}
q_{j|\br} = 1-p_{j|\br}.
\end{equation}
Together with \eqref{intpow}, the fraction of remaining interference becomes
\begin{equation}
v_j = 1 - \expect \limits_{\br} \left(1 -p_{j|\br}\right)^2.
\end{equation}
In order to implement \eqref{softcanrule}, we need to know $p_{j|\br}$, the error probability within user group ${\cal G}_j$ given the receive word $\br$.

Since we do not know how to calculate $p_{j|\br}$, we will use upper and lower bounds on the fraction of remaining interference.
For the upper bound, we base our soft-cancellation on $p_{j|\tr_{1:}}$ instead of $p_{j|\br}$.
This yields
\newcommand{\vju}{{v}_j^{\text u}}
\newcommand{\vjl}{{v}_j^{\text l}}
\begin{align}
v_j < \vju &=  1 - \int\limits_{\setR}  
\frac{
{\text Q}\left(-\frac x{\sqrt{I}} \right)^{2^{K}-1}{\text e}^{-\frac{(x-\sqrt{NP_j})^2}{2I}}
}
{[1+1/{\text F}(x)] \sqrt{2\pi I}}
{\text d}x\\
&=1 - \int\limits_{\setR}  
\frac{
{\text Q}\left(x-\sqrt{\eta NP_j} \right)^{2^{K}-1}
}
{1+1/{\text F}\left(\sqrt{\eta NP_j}-x\right) }
{\text D}x
.
\label{vu1}
\end{align} 
For the lower bound, we assume perfect knowledge of whether a decision is correct or not. 
This implies
\begin{equation}
\label{vlower}
v_j > \vjl =  p_j.
\end{equation}
In the sequel, we will refer to these bounds when addressing the performance of decision-directed soft-cancellation.

\section{Finite User Case}
\label{finiteM}

For a finite number of users, the interference does become correlated during iterations.
This is a severe problem for practical algorithms, as well as for computer simulations that shall support the calculations in the previous sections.
This problem is often addressed by means of approximate message passing \cite{opper:05,donoho:09} and its various recent improvements \cite{ma:17,rangan:19,takeuchi:20}. Due to the multidimensional nature of the codebook, approximate message passing is anything, but straightforward to apply to the problem at hand and is left for future work.
In the sequel, we propose a simple low cost alternative that, in Section~\ref{numres}, turns out to work, though also leaving some room for further improvement. 

In order to cancel interference, an estimate for the interfering signal due to group ${\cal G}_j$ is calculated for all groups.
The estimate is formed by
\begin{equation}
\label{estint}
\bi_j = \sqrt s_j \sum\limits_{m\in{\cal G}_j} \bch_m (1-p_{m})
\end{equation}
with $p_m$ denoting the estimated probability that the decision for codeword $\bch_m$ is incorrect and $s_j$ being a scale factor that will be discussed in the sequel.
For $s_j=1$, this is the cancellation rule discussed in Section~\ref{intcan}.

If all codewords in \eqref{estint} were orthogonal, the total interference power would be given by
\begin{equation}
\expect ||\bi_j||^2 = \frac{s_j  P_j}M \sum\limits_{m\in{\cal G}_j} (1-p_m)^2, 
\end{equation}
since $\expect ||\bch_m||^2 = P_j/M$ for users in group ${\cal G}_j$. Since the codewords are not orthogonal for finite number of users $M$, the estimated interference is typically larger. This overestimation leads to a too aggressive interference cancellation policy which is prone to error propagation. To avoid such harm, we set
\begin{equation}
s_j = \frac{\frac {P_j}M \sum\limits_{m\in{\cal G}_j} (1-p_m)^2  }
{ \left|\left|\sum\limits_{m\in{\cal G}_j} \bch_m (1-p_{m}) \right|\right|^2}.
\end{equation}
An additional minor improvement is achieved, if the re-normalization is repeated among user groups.
The total estimate of interference is, thus, formed as
\begin{equation}
\bi = \sqrt s \sum\limits_{j=1}^J \bi_j
\end{equation}
with
\begin{equation}
s = \frac{\sum\limits_{j=1}^J \alpha_j v_j P_j}
{\left|\left|\sum\limits_{j=1}^J \bi_j\right|\right|^2}.
\end{equation}
These two re-normalizations of the interference estimate strongly improve the block error rate simulated in Section~\ref{numres}.
\section{Improving Convergence}
\label{imcon}
Irregularity aids the convergence of iterative systems.
This phenomenon is well studied, e.g.\ in the context of low-density parity check codes \cite{mackay:03}. It has also been observed for iterative multiuser decoding in \cite{caire:01b}.

There are various way to introduce irregularity into iterative multiuser decoding. In the sequel, we will address power imbalances among users.

While for low rates, equal power levels for all users turn out optimal, this does not hold if the rate exceeds some finite threshold.
This effect was first observed in \cite{caire:01b}. In the sequel, we apply the ideas of power optimization laid out in \cite{caire:01b} to Gaussian random coding assuming an infinite number of users.


Power 
optimization can be performed by linear programming.
This is possible, as the multiuser efficiency is identical for all user groups.
Its evolution during iterations can be tracked by the dynamical system defined in \eqref{etadyn} and \eqref{etavmap}.
The mappings from the multiuser efficiency to the fractions of remaining interference depend on the particular way, interference cancellation is implemented.
For the upper and lower bounds considered in this paper, they can be found in \eqref{vu1} and \eqref{vlower} via \eqref{subf73}.

In order for iterations to converge, we need to ensure that the multiuser efficiency at the next iteration exceeds the current multiuser efficiency by an arbitrarily small margin $\epsilon>0$.
This can be ensured by the linear program
\begin{equation}
\left\{
\begin{array}{ll}
\min\limits_{\alpha_1,\dots,\alpha_J}  &\sum\limits_{j=1}^J \alpha_j  P_j \\
\mbox{subject to} & \alpha_j \ge 0 \qquad \forall j\\
&\sum\limits_{j=1}^J \alpha_j P_j v_j(\eta) < \frac1{\eta+\epsilon} - 1\qquad  \forall \eta \in {\cal E}\\
&\sum\limits_{j=1}^J {\alpha_j} = 1
\end{array}
\right..
\label{linprog}
\end{equation}
for an appropriately chosen interval ${\cal E} \subset [0;1]$.
Its lower end may be chosen as large as the multiuser efficiency before the first iteration.
Its upper end determines the error probability after iterations have converged. It is a design parameter of the multiuser system. So is the margin $\epsilon$. The smaller it is, the more iterations are needed.

The 
powers $P_j$
are quantized versions of the optimal 
distribution of powers. 
The larger the number of groups $J$, the better is the approximation to the optimal 
distribution. 
This indirect way of power optimization is chosen, as the function $v_j(\eta)$ depends in a non-convex way on the powers of the users, see \eqref{vu1}, but not on the group size.

\section{The Near-Far Gain}
\label{nearfar}

In practice, receive powers of users will vary anyway due to different propagation conditions among users. This can be utilized to reduce the average transmit energy per bit following the ideas of \cite{caire:07}, see also \cite[Chapt.~5]{mueller:99} and \cite{tse:98a}. A similar concept was popularized more recently under the generic term {\it non-orthogonal multiple-access (NOMA)} \cite{ding:17}. 
In context of the current work, one simply needs to adjust the weights $w_j$ in the objective function of \eqref{linprog}.
 
The origin of the near-far gain is sometimes obscured in recent papers on NOMA. 
In fact, the near-far gain is difficult to understand intuitively, if one is too focussed on a direct boost in data rate. Information theory, however, establishes a fundamental duality between data rate and energy per bit. If we aim to minimizing the energy per bit for a given target data rate instead, the near-far gain is very intuitive.

For iterative decoding, in general, and successive cancellation, in particular, to work close to capacity limits, irregularity is required. This irregularity can be provided by the system design at some price, e.g., protecting some data symbols by more parity-checks than others. This comes at the expense of more redundancy and, thus, reduced data rate.
In successive cancellation, the equivalent is larger transmit power. Here the price is paid in dual currency: in the energy per bit.

Near-far situations provide irregularity for free. It takes the form of receive power imbalances.
These natural receive power imbalances are not exactly distributed as they are supposed to be. Adjustment is needed.
However, it is less effort to adjust from already imbalanced receive powers than starting from the worst case: equal received powers. The reduced adjustment effort is the near-far gain measured in reduced transmitted energy-per bit. It may be quantified running the linear program \eqref{linprog} once with unit weights and once with weights provided by natural attenuation, then comparing the two total powers \eqref{totP}. Standard methods 
can be applied for currency conversion into bits/s/Hz.

The near-far gain is not restricted to path loss alone.
Long-term fading typically exhibits dynamics slow enough to be utilized in the same or a similar way. Given the system settings, even short-term fading can be utilized. These details have been extensively discussed in the recent NOMA literature, see, e.g., \cite{ding:17} for a survey.

\section{Numerical Results}
\label{numres}
Numerical results can be difficult to obtain.
If the number of bits per user exceeds values around 35, the exponent $2^{K}-1$ in various equations becomes numerically unstable to evaluate, as the basis is very close to unit.
This can be circumvented as follows:
\begin{equation}
{\text Q}(x)^a = {\text e}^{a \ln (1-{\text Q}(-x))} = \prod\limits_{i=1}^\infty {\text e}^{-a {\text Q}(-x)^i/i}
\end{equation}
For sufficiently large $a$, all factors for $i>1$ are so close to unity that they can be ignored.
Furthermore, the Gaussian integration can be tedious. We recommend Gauss-Hermite quadrature with several hundred terms (we used 300).


\subsection{Equal Path Loss for All Users}

\begin{figure}
\epsfig{file=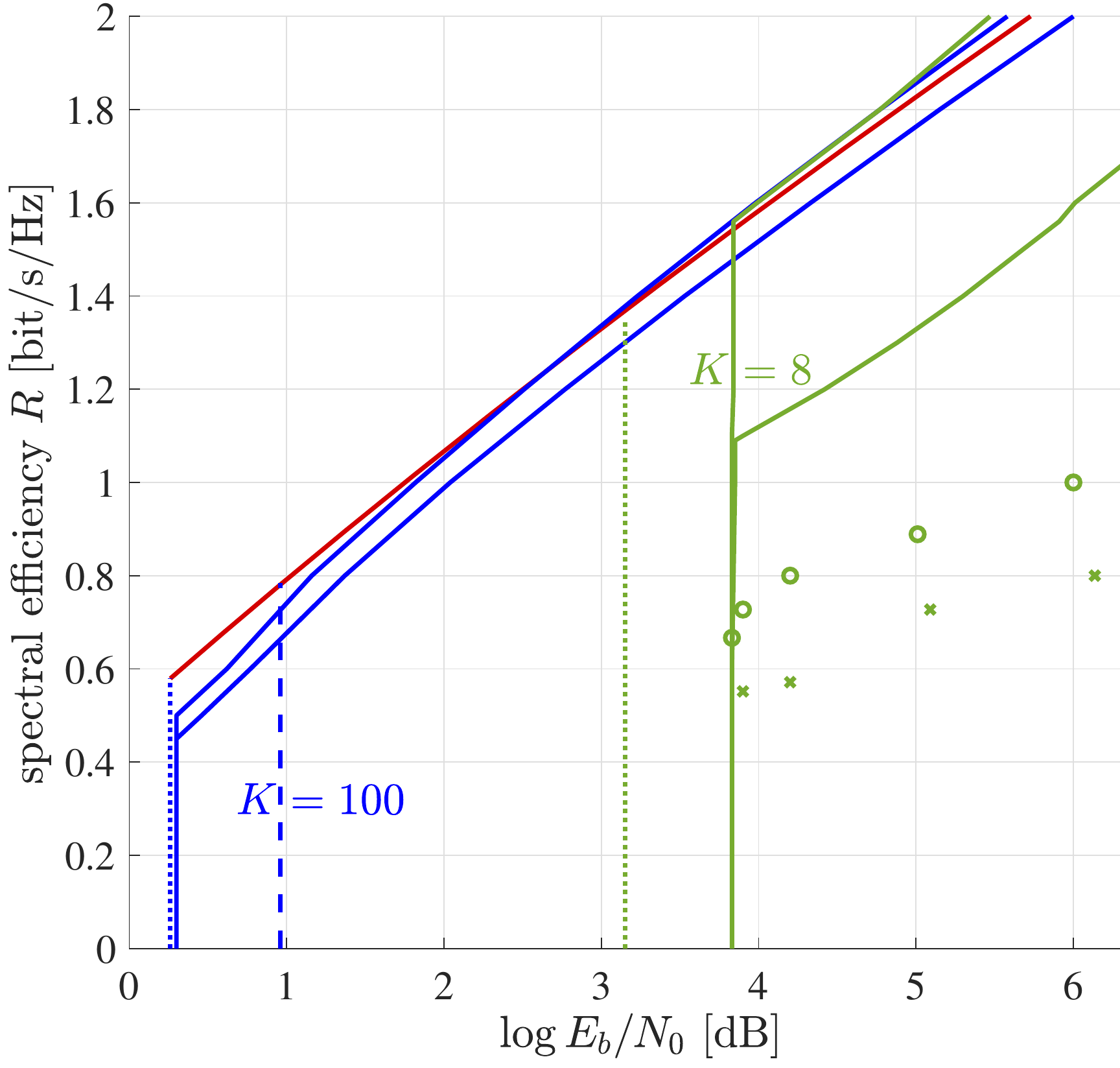,width=\columnwidth}
\caption{Spectral efficiency vs.\ rate-compensated signal-to-noise ratio for per user block error rate $10^{-3}$. The solid lines refer to our inner and outer bounds introduced in Section~\ref{intcan}. The dashed and dotted lines refer to the best inner and outer bounds of \cite{zadik:19}. The two indistinguishable red lines are given by setting $w_\ell=1$ in \eqref{outbou} for $K=100$ and $K=8$. Points marked by circles and crosses refer to simulations with 256 and 32 users in the largest group, respectively.
\label{zpt}}
\end{figure}

Fig.~\ref{zpt}
shows the trade-off between spectral efficiency and power efficiency for block error rate $10^{-3}$ and power distribution optimized among users with parameter $\epsilon = 10^{-3}$, suitable\footnote{The choice of ${\cal E}$ is not totally trivial. The lower end can be chosen arbitrarily close to 0, but eventually also somewhat larger to speed up the linear program. The choice of the upper end determines the final error probability by means of a strictly monotonous function (more remaining interference implies higher error probability). Here, the method of interval nesting was applied. Typical values for the upper end for target error probability $10^{-3}$ range from 0.95 to 0.99 depending on the signal-to-noise ratio.} choice of ${\cal E}$, and equal message lengths for all users.
 
There are two paradigms: the equal power regime and the distributed power regime.

\subsubsection{Equal Power Regime}
In the equal power regime, all users transmit at the same power. In this regime, our outer and inner bounds coincide and spectral efficiency is independent of power efficiency. Iterations proceed until the multiuser efficiency becomes very close to unity and nearly all interference has been removed. Thus, the error probability relates to  $E_{\text b}/N_0$ approximately as
\begin{equation}
\label{ebnomin}
P_{\text e} = 1- \int\limits_{\setR} {\text Q}\left(x - \sqrt{2K\frac{E_{\text b}}{N_0}} \right) ^{2^{K}-1} {\text D}x.
\end{equation}
In this regime, the error probability is determined by the minimum required $E_{\text b}/N_0$ for given amount of information bits per user.

\subsubsection{Distributed Power Regime}
In the distributed power regime, the sizes of the user groups are optimized by the linear program \eqref{linprog}. Within each group, the power per user is the same, but it differs from group to group. In order to reduce granularity effects of the discretization of the power distribution, the linear program is run with more than hundred power groups. However, the linear program returns most of them empty (without users). This indicates that the the optimum number of groups is finite. The larger the signal-to-noise ratio, the larger is the optimal number of groups. For the minimal signal-to-noise ratio, all users are in the same group. Any larger signal-to-noise ratio has its own individually optimal power distribution. All these observations are in line with the results on power optimization in iterative decoding of convolutionally encoded code division-multiple access reported in \cite{caire:04}. In this reference, some optimal power distributions are shown. They are qualitatively very similar to the ones found in this work.

For large values of spectral efficiency, the outer bound of \cite{zadik:19} (red line in Fig.~\ref{zpt}) becomes tighter than our outer bound which results from the genie-added lower bound on the remaining interference \eqref{vlower}. 
For $K=100$, inner bound and best outer bound differ by about a quarter of a decibel, while for $K=8$, they differ by approximately 1.5 dB.

\subsubsection{Finite Number of Users}
Simulations for finite number of users utilizing double re-normalization according to Section~\ref{finiteM} are shown as circles and crosses. For all simulation points 25000 symbols are transmitted and up to 30 iterations are performed. For the fives simulation points around $\log E_{\text b}/N_0=4$ dB a single group of users was used with $M=256$ and $M=32$ for the circles and the crosses, respectively. Performance strongly increases with the number of users, as the codewords become more and more orthogonal\footnote{Simulations with larger number of users were not feasible on the author's computer due to lack of memory.}. For the simulation points at 5 dB and above, the power profile is optimized by try and error, as the linear program \eqref{linprog} cannot be utilized. To keep the size of the search space reasonable, only $J=2$ groups are considered. In all cases, the largest group of users contains $256$ and $32$ users, respectively, that operate with minimum power to achieve the target block error rate of $10^{-3}$. At average $\log E_{\text b}/N_0 \approx 5$ dB, a second group with 64, respectively 8, users of larger power is added. This increases the total number of users to 320, respectively 40. Due to the users with higher power, the average $E_{\text b}/N_0$ raises. At the same time, the parameter $N$ can be reduced, such that the spectral efficiency increases, as well. At average $\log E_{\text b}/N_0 \approx 6$ dB, the second user group is chosen twice as large as for $\log E_{\text b}/N_0 \approx 5$ dB.
Although the simulation results fall quantitatively behind the theoretical predictions for $M\to\infty$ due to the lack of orthogonality between codewords, they show the same qualitative behavior as the proposed theory.
However, recent subsequent works \cite{mohammadkarimi:21,hsieh:21} show that simulations based on approximate message passing instead of basic soft interference cancellation perform well between the asymptotic bounds proposed in this paper.

\subsubsection{Minimum Signal-to-Noise Ratio}
\begin{figure}
\epsfig{file=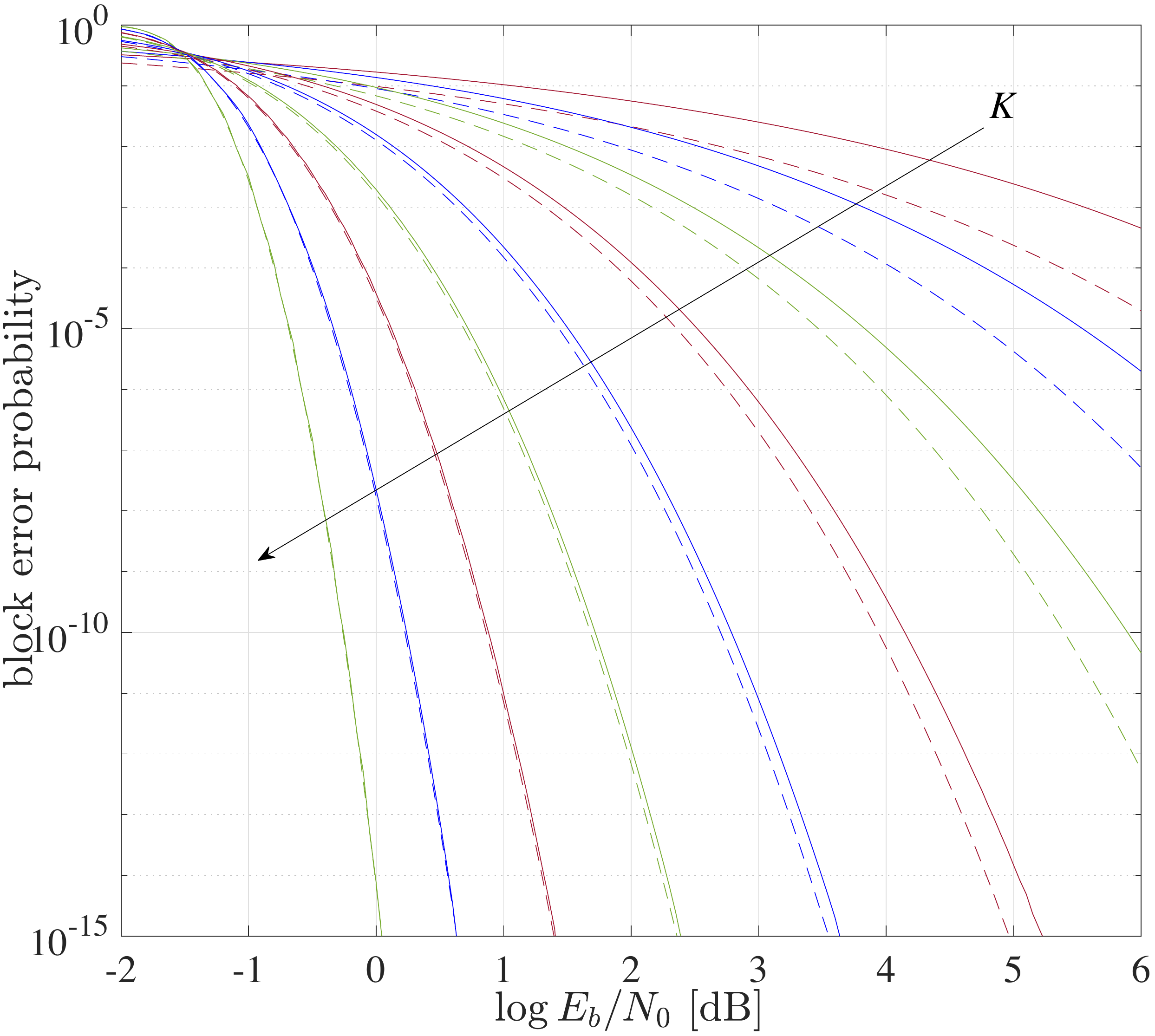,width=\columnwidth}
\caption{Block error probability at minimum required $E_{\text b}/N_0$ for various message lengths $K=4,8,\dots, 512, 1024$ (following arrow). Solid and dashed lines refer to \eqref{ebnomin} and \eqref{ebnominb}, resp.
\label{peebno}}
\end{figure}

The block error probability at the minimum possible $E_{\text b}/N_0$ is shown in Fig.~\ref{peebno}
for various message lengths $K$. The solid and dashed lines refer to \eqref{ebnomin} and the lower bound \cite{polyanskiy:11}
\begin{equation}
\label{ebnominb}
P_{\text e} > 1- {\text Q}\left({\text Q}^{-1}\left(2^{-K}\right) - \sqrt{2K\frac{E_{\text b}}{N_0}} \right)
\end{equation}
respectively.
While the lower bound is tight for long messages, it may be loose by several orders of magnitude for short messages. The looseness for $K=8$ can also be observed in Fig.~\ref{zpt}.

\subsection{Discretized Path Loss Model}

Path loss is commonly modeled by a continuous statistical distribution. The linear program \eqref{linprog}, however, can only handle a finite number of different received power levels.
Therefore, we use a simple discretized model, in the sequel.

Let there only be $L$ different fading weights $\sqrt w_1,\dots,\sqrt w_L$.
Partition each of the $J$ user groups into $L$ subgroups with the $\ell^{\text{th}}$ subgroup experiencing fading gain $\sqrt w_\ell$ and $\alpha_{j\ell}$ denoting the fraction of users in the $\ell^{\text{th}}$ subgroup of group ${\cal G}_j$.
We modify the linear programm \eqref{linprog} to read
\begin{equation}
\left\{
\begin{array}{cl}
\min\limits_{\alpha_{j\ell},\forall j,\ell} &\sum\limits_{j=1}^J\sum\limits_{\ell=1}^L \alpha_{j\ell} w_\ell P_j\\
\mbox{s.t.} & \alpha_{j\ell} \ge 0 \qquad \forall j,\ell\\
&\sum\limits_{j=1}^J \sum\limits_{\ell=1}^L\alpha_{j\ell} w_\ell P_j v_j(w_\ell \eta) < \frac1{\eta+\epsilon} - 1\quad  \forall \eta \in {\cal E}\\
&\sum\limits_{j=1}^J \alpha_{j\ell} = \Pr(w_\ell) \qquad \forall \ell 
\end{array}
\right.
\end{equation}
where we introduced additional constraints
to prevent the linear program from changing the distribution of the fading gains.

\begin{figure}
\epsfig{file=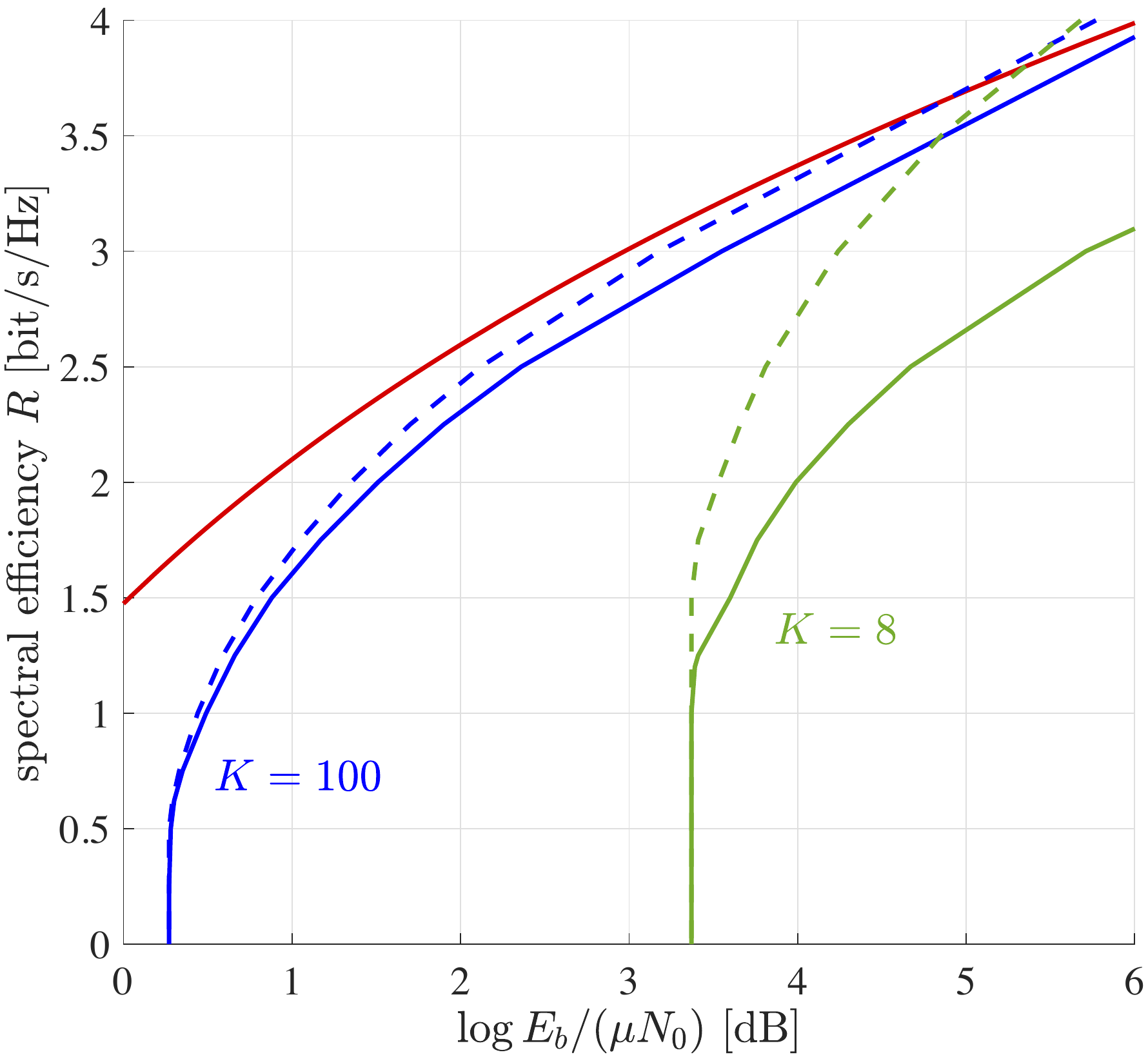,width=\columnwidth}
\caption{Spectral efficiency vs.\ rate-compensated transmit signal-to-noise ratio for per user block error rate $10^{-3}$. The two indistinguishable red lines are the outer bounds \eqref{outbou} for $K=100$ and $K=8$.  All curves for $L=10$.
The other lines refer to the inner and outer bounds introduced in Section~\ref{intcan}. 
\label{zptnf}}
\end{figure}

Considering a linear path loss model and free space propagation (which gives similar results as a circular path loss model with attenuation exponent 4), we set the fading weights to
\begin{equation}
\sqrt w_\ell = \frac 1{\ell}
\end{equation}
and denote the average fading gain by
\begin{equation}
\mu = \frac 1L \sum\limits_{\ell=1}^L w_\ell.
\end{equation}
We redo the numerics of Fig.~\ref{zpt} under otherwise identical conditions.
However, we measure power efficiency in transmitted energy per bit normalized to the average fading gain, i.e.\ $E_{\text b}/(\mu N_0)$, which obeys the upper bound \cite[Eq.~(5.24)]{mueller:99}
\begin{equation}
\label{outbou}
\frac{E_{\text b}}{\mu N_0} \ge 
\frac 1{2R} \sum\limits_{\ell=1}^L
{w_\ell}  \left[{
4^{aR\ell/L} - 4^{aR(\ell-1)/L}
}\right]. 
\end{equation}
Here,
\begin{equation}
a = 1-P_{\text e} - H_2(P_{\text e})/K.
\end{equation}
is a correction factor accounting for finite blocklength, see \cite{zadik:19} for details.
Numerical results are shown in
Fig.~\ref{zptnf}.
In contrast to Fig.~\ref{zpt}, there is no sharp transition between the equal and the distributed power regime. The gap between our two bounds has widened. 

The equal power regime has moved towards lower values of ${E_{\text b}}/{(\mu N_0)}$. The effect is particularly pronounced for short message lengths, cf.~$K=8$.
This happens, as there is no side constraint enforcing fairness among users: While the overall block error probability is still $10^{-3}$, users in bad channel conditions experience larger error probability. Users in good channel conditions compensate for that. For users in good channel conditions, low error probability is very cheap in terms of transmit power. As a result, this overcompensates the excess power required by users in bad channel conditions.

\section{Conclusions}
\label{conc}
Random codes perform well for very massive multiple-access even if users have short messages. They can be iteratively decoded by soft-cancellation of interference, but may required power optimization to create enough irregularity to allow iterations to converge.

In the large user limit, simplex constellations in $2^8$ dimensions carrying 8 information bits are  hardly more than 1.5 dB behind random codes of infinite length, if spectral efficiency is larger than 1.1 bits/s/Hz. This gap is the larger, the smaller is the number of users. Further research into iterative algorithms for soft cancellation, e.g., utilizing ideas of approximate message passing, may turn out helpful.

For high spectral efficiency, users should be received at unequal power levels. This is beneficial in practice, as wireless propagation conditions unavoidably create such power imbalances.

 \appendices

 \section{Unconditional Block Error Probability}
 \label{appproof}
 
 Let $\bz\in\setR^{NM}$ denote the vector of interference and noise. 
The ensemble-averaged block error probability for any user in group ${\cal G}_j$ given the Euclidean norms of receive word $\br$ and interference-and-noise vector $\bz$, $r=||\br||$ and $z=||\bz||$, respectively, is given by \cite{mueller:20d}
\begin{equation}
p_{j|r,z} = 1- {\text Q}_{\frac{MN}2}\left(\frac{r}{\sqrt {P_j/M}},\frac{z}{\sqrt {P_j/M}}\right)^{2^{K}-1}
\end{equation}
with ${\text Q_a(b,c)}$ denoting the generalized Marcum Q-function.
Although the Euclidean norms of received word and interference-and-noise vector are not independent of each other, they can be constructed out of three statistically independent random variables $\chi$, $\zeta$, and $\gamma$ \cite{mueller:20d} by
\begin{align}
z ^2 &= \chi^2 + \zeta^2\\
r^2 & = \chi^2 + (\zeta + \gamma)^2.
\end{align}
As discussed in \cite{mueller:20d}, $\zeta$ and $|\chi|$ are the radial and the Euclidean norm of the tangential component of noise and interference, respectively.
Furthermore, $|\gamma|$ is the Euclidean norm of the transmitted codeword.
Thus, $\zeta$ is zero mean Gaussian with variance $I$, $\gamma^2M/P_j$ and $\chi^2/I$ are chi-square distributed with $MN$ and $MN-1$ degrees of freedom, respectively.

The conditional error probability can be written as 
\begin{equation}
p_{j|\chi,\zeta,\gamma}  = 1- {\text Q}_{\frac {MN}2}\left( \scriptstyle \sqrt{\frac{\chi^2 + (\zeta+\gamma)^2}{P_j/M}},\sqrt{\frac{\chi^2 +\zeta^2}{P_j/M}}  \right)^{2^{K}-1}
\label{pmac}.
\end{equation}
Both arguments of the generalized Marcum Q-function in \eqref{pmac} linearly scale with $M$. 
The term $(\chi^2 +\zeta^2)/I$ is chi-square distributed with $MN$ degrees of freedom. Its mean and standard deviation are $MN$ and $\sqrt{2MN}$, respectively. 
Its distribution, if normalized by $M$, converges to a mass point at $N$. 
Due to the term $P_j/M$ in the denominator, the second argument of the generalized Marcum Q-function asymptotically scales linearly in $M$.
The first argument is even slightly larger due to the addition of $\gamma$. However, $\gamma$ does not scale with the number of users, so asymptotically both terms scale in the same way.
Thus, we are interested in the behavior of the generalized Marcum Q-function when all arguments grow over all bounds.
In Appendix~\ref{appc}, we show
\begin{equation}
\label{defmarcum}
\lim\limits_{M^\prime\to\infty}{\text Q}_{aM^\prime}(M^\prime-\epsilon,M^\prime) = {\text Q}(\epsilon-a)
\end{equation}
with ${\text Q}(\cdot)$ denoting the standard Gaussian Q-function.
\newcommand{\ase}{\stackrel{\cdot}=}
Thus, we obtain
\begin{align}
\label{qroots}
p_{j|\chi,\zeta,\gamma}
&\ase 1- {\text Q}\left( \scriptstyle \sqrt{\frac{\chi^2 +\zeta^2}{P_j/M}}  - \sqrt{\frac{\chi^2 + (\zeta+\gamma)^2}{P_j/M}} - \frac{N\sqrt{MP_j}}{2\sqrt{\chi^2+\zeta^2}} \right) ^{2^{K}-1}
\end{align}
with $\ase$ denoting asymptotic equivalence for $M\to\infty$.
With probability approaching 1 for large $M$, we have
\begin{equation}
\chi^2 \gg \zeta^2 \qquad \wedge  \qquad \chi^2 \gg (\zeta + \gamma)^2.
\end{equation}
Thus, we can develop the roots in \eqref{qroots} into first order Taylor series at $\chi^2$ and obtain
\begin{align}
p_{j|\chi,\zeta,\gamma}
&\ase 1-  {\text Q}\left( \frac{-NP_j - \gamma^2 -2|\gamma| \zeta}{2|\chi| \sqrt {P_j/M}} \right)^{2^{K}-1}
\end{align}
The random variable $|\gamma|\sqrt{M/P_j}$ is chi-distributed. Thus, its variance is upper bounded by $\frac12$. This implies that the variance of $|\gamma|$ vanishes for large $M$. This is in contrast to $\gamma^2$ and $\zeta$ which have variance $2NP_j$ and $I$ given in \eqref{defI}, respectively.
For $M\to\infty$, $|\gamma|$ is arbitrarily closely approximated by its asymptotic mean $\sqrt{NP_j}$. 
Similar considerations imply that $|\chi|$ may be replaced by its asymptotic mean $\sqrt{IMN}$.
This gives
\begin{align}
p_{j|\chi,\zeta,\gamma}&\ase p_{j|\zeta,\gamma}\\
&\ase 1- {\text Q}\left( \frac{-NP_j - \gamma^2 -2\sqrt{NP_j} \zeta}{2\sqrt{INP_j} } \right) ^{2^{K}-1}
\end{align}
The argument of the Q-function is the sum of a constant and two random variables with asymptotic distributions
\begin{align}
-\frac{\zeta}{\sqrt I}  & \sim {\cal N} (0,1)\\
 -\frac{\gamma^2}{2\sqrt{INP_j}} & \sim {\cal N} \left(-\frac{\sqrt{NP_j}}{2\sqrt I}, \frac{1}{2IM}\right).
\end{align}
The second random variable turns into a constant as $M\to\infty$.
This implies
\begin{align}
p_{j|\zeta,\gamma} &\ase p_{j|\zeta}\\
&\ase 1- {\text Q}\left( \frac{-\sqrt{NP_j}-\zeta  }{\sqrt{I} } \right) ^{2^{K}-1}.
\label{pjzeta}
\end{align}

From the three random variables $\chi$, $\zeta$, and $\gamma$, only $\zeta$ has survived the infinite user limit. The variance of $\gamma$ has vanished. The variance of $\chi$ has not vanished, but the influence of $\chi$ on the conditional block  error probability has done so.
It can be seen from \cite{mueller:20d} that $\zeta$ is the radial component of noise and interference relative to the true codeword.
Averaging over the Gaussian random variable $\zeta$, we obtain \eqref{subf73}.

 \section{Limit of the Generalized Marcum Q-Function}
 \label{appc}
The noncentral chi-square distribution with $k$ degrees of freedom and non-centrality parameter $\lambda$ follows the CDF
\begin{equation}
1-{\text Q}_{\frac k2}\left(\sqrt\lambda,\sqrt x\right) \to 1- {\text Q}\left(\frac{x-\mu}{\sigma}\right)
\end{equation}
which converges to the Gaussian distribution of same mean $\mu$ and variance $\sigma^2$ due to the central limit theorem.
We have
\begin{align}
\label{mstd}
\mu &= k+\lambda, \qquad
\sigma^2 = 2k + 4\lambda
\end{align}
Letting $k=2aM$, $\lambda=(M-\epsilon)^2$, and $x=M^2$, we get
\begin{align}
{\text Q}_{aM}(M-\epsilon,M) &\to{\text Q}\left(
\frac{x-k-\lambda}{\sqrt{2k+4\lambda}}
\right)\\
 &={\text Q}\left(
\frac{M^2-2aM-(M-\epsilon)^2}{\sqrt{4aM+4(M-\epsilon)^2}}
\right)
\end{align}
which for $M\to\infty$ converges to \eqref{defmarcum}.
\bibliography{lit}
\bibliographystyle{IEEEtran}

\end{document}